\documentclass[12pt, draftclsnofoot, onecolumn]{IEEEtran}

\usepackage{cite}	
\usepackage{graphicx} 
\DeclareGraphicsExtensions{.pdf,.eps}
\usepackage[latin1]{inputenc} 
\usepackage[T1]{fontenc} 
\usepackage{amsmath,amsfonts,amsbsy,amssymb} 
\usepackage{mathabx} 
\usepackage{amssymb} 
\usepackage{amsmath}
\usepackage{amsthm}	
\usepackage{mathrsfs}
\usepackage[nolist]{acronym} 
\usepackage{tabularx} 
\usepackage{multirow}
\usepackage{wasysym}
\usepackage{float}
\usepackage{color} 
\usepackage{algorithm}
\usepackage[noend]{algpseudocode}

\usepackage{enumitem} 

\newtheorem{lemma}{Lemma}
\newtheorem{theorem}{Theorem}

\newtheorem{Problem}{Problem}

\usepackage[colorlinks,bookmarksopen,bookmarksnumbered,linkcolor=black,citecolor=black,urlcolor=black]{hyperref}

\begin{document}
	
	\title{Ultra Reliable Communication via Optimum Power Allocation for HARQ Retransmission Schemes}
	\author{Endrit Dosti, Mohammad Shehab, Hirley Alves and Matti Latva-aho\\

		\thanks{E. Dosti is with the Department of Signal Processing and Acoustics, Aalto University, Finland. Contact: firstname.lastname@aalto.fi 
		
		M. Shehab, H. Alves and M. Latva-aho are with the Centre for Wireless Communications (CWC), University of Oulu, Finland. Contact: firstname.lastname@oulu.fi.}
		\thanks{This work is partially supported by Academy of Finland 6Genesis Flagship (Grant no. 318927), Aka Project EE-IoT (Grant no. 319008).}
	}
	%
	\maketitle
	
	\begin{abstract}
		
		In this work, we develop low complexity, optimal power allocation algorithms that would allow ultra reliable operation at any outage probability target with minimum power consumption in the finite blocklength regime by utilizing Karush-Kuhn-Tucker (KKT) conditions. In our setup, we assume that the transmitter does not know the channel state information (CSI). First, we show that achieving a very low packet outage probability by using an open loop setup requires extremely high power consumption. Thus, we resort to retransmission schemes as a solution, namely Automatic Repeat Request (ARQ), Chase Combining Hybrid ARQ (CC-HARQ) and Incremental Redundancy (IR) HARQ. Countrary to classical approaches, where it is optimal to allocate equal power with each transmission, we show that for operation in the ultra reliable regime (URR), the optimal strategy suggests transmission with incremental power in each round. Numerically, we evaluate the power gains of the proposed protocol. We show that the best power saving is given by IR-HARQ protocol. Further, we show that when compared to the one shot transmission, these protocols enable large average and maximum power gains. Finally, we show that the larger the number of transmissions is, the larger power gains will be attained.
		
	\end{abstract}
	
	\section{Introduction}
	\label{Introduction}
	Modern communication systems play an essential role in everyone's daily life. Throughout the history of the development of these systems, the goal has been to enable communication with higher data rates. This trend is expected to be continued in the future as well. However, in the next generation systems, the vision will be to connect all devices that benefit from an Internet connection to create a data driven society and pave the road towards 6G \cite{6g}, thus resulting in the creation of the Internet of Things (IoT). A key characteristic of the IoT is that most of the wireless connections are expected to be generated by the autonomous devices rather than by the human-operated terminals. To successfully implement this vision, wireless communication systems will have to support a much larger number of connected devices, and at the same time fulfill much more stringent requirements on latency and reliability that what current standards can guarantee \cite{Durisi_1}.
	
	For this purpose, 5G introduced at least two new operating modes. The first one is Massive Machine-to-Machine Communications, which is related to designing a wireless system that can support a large number of simultaneously connected devices (\textit{e.g.} more than 10s of thousands). This will pave path for seamless and ubiquitous connectivity of heterogenous devices \cite{Palattella,Nokia}. The second operating mode is Ultra Reliable Low Latency Communications (URLLC). This mode refers to insuring a certain level of communication service almost 100\% of the time, while satisfying very stringent delay requirements (\textit{i.e.} in the order of 4 ms). \cite{Popovski2014}. URC is essential in many applications such as reliable wireless coordination among vehicles, reliable cloud connectivity, critical connections for industrial automation \cite{Yilmaz,MTC_6g}. 
	
	To cope with the presence fading in wireless comunication systems, several diversity schemes are implemented in order to create redundancy by transmitting the signal through several independent fading paths and then combining it accordingly at the receiver. This provides higher reliability at the cost of increased power consumption. In this context, the problem of optimum power allocation in URC scenarios has recently attracted lots of attention \cite{TVT_Shehab,Onel,EEE}. For example, in \cite{Onel} the authors desigend an optimum power allocation algorithm to maximize the energy efficiency in Single-Input Multiple-Output (SIMO) systems with stringent reliability constraints. Moreover, the utilization of different retransmission schemes is a classical approach to provide larger diversity gains and achieve ultra-relibability. The most popular schemes are those which implement repetitive or parallel retransmission techniques \cite{4524270, 7349145, 7045615, 7452354, 6949093, WiMAX, 3GPP,Larsson1, Chaitanya2016, Tumula2, 5754756, me, me1, Kim2013, Devassy2014, Makki2014, Makki2015, Durisi_HARQ}. In the former, the transmitter sends the same codewords in all the possible fading paths. While in the latter, the transmitter utilizes different and jointly designed codewords to construct the packets.
	
	The simplest retransmission protocol which utilizes repetition coding, is Automatic Repeat Request (ARQ) \cite{4524270}, where transmitter sends the packets until it receives an acknowledgment (ACK) from the receiver, or until the maximum allowed number of retransmissions is exhausted. A more robust retransmission protocol family that utilizes repetition coding is Hybrid-ARQ (HARQ) \cite{7349145}. The difference is that in this family the receiver buffers all the packets and utilizes them to correctly decode the information. HARQ protocols are classified based on how the receiver combines the packets. For instance, when the receiver makes decisions based on the selected packet with the highest Signal-to-Noise Ratio (SNR), the scheme is called Selection Combining (SC) HARQ \cite{7045615}. Secondly in Chase Combining (CC) HARQ \cite{7452354}, the receiver can do Maximal Ratio Combining (MRC) of the received signals and extract the information. All these schemes boost reliability at the cost of increased latency and comlexity of the receiver design. In addition, several retransmission protocols utilize parallel coding schemes, where they utilize Incremental Redundancy (IR) \cite{6949093}. Therein, the transmitter sends new information with each retransmission. This is achieved by splitting the parent codeword into several sub-codewords. The utilization of this family of retransmission protocols has been embraced by several systems, such as Worldwide Interoperability for Microwave Access (WiMAX), Long-Term Evolution (3GPP LTE), and 5G NR \cite{WiMAX, 3GPP}. 
	
	\subsection{Related Work}	
	The problem of adaptive and optimal resource allocation in diversity and retransmission schemes has been investigated in several papers \cite{Larsson1, Chaitanya2016, Tumula2, 5754756, me, me1, Kim2013, Devassy2014, Makki2014, Makki2015, Onel, Durisi_HARQ}. Mainly, the optimization parameters are different performance metrics, such as outage probability, throughput, effective capacity, delay, power/energy efficiency etc. For instance, In \cite{Larsson1} the authors derive closed form expressions for the rate-maximized throughput of ARQ for independent distributed Nakagami-$m$ block fading interfering channels. The authors of \cite{Durisi_HARQ} studied the trade-off between short paket transmission and utlizing HARQ in low latency communication. They showed that HARQ may significantly outperform finite blocklength for a given set of latency, reliability and bit count values. Energy-efficient adaptive power allocation for
	three incremental multiple-input-multiple-output (IMIMO) systems employing ARQ, CC-HARQ and IR-HARQ are considered in \cite{Chaitanya2016}. There, the authors formulate and provide closed form solutions for the proposed geometric programming	problem (GPP) to minimize the rate outage probability. However, they did not consider the minimum power consumption problem which we target in our analysis. 
	
	An issue that has gathered much attention from the research community is power allocation between different retransmission rounds. In \cite{Tumula2} the authors provide a closed form approximation for the outage probability for CC-HARQ protocol. Furthermore, they formulate and solve the power allocation problem as a GPP. A limited power allocation strategy valid for a maximum of two transmissions for IR-HARQ protocol is proposed in \cite{5754756}. In contrast to the work presented there, we derive closed form expression for the outage probability of IR-HARQ. Furthermore, we obtain optimal power allocation algorithms of low complexity for any number of transmissions and prove analytically that the proposed solutions are globally optimal. Also, unlike \cite{Tumula2, 5754756} we provide comparisons between the performance of proposed power allocation algorithms for the three most popular retransmission protocols, respectively ARQ, CC-HARQ and IR-HARQ. 
	
	Moreover, the analysis in all the above mentioned papers is done under the assumption of asymptotically long codewords, where the length of metadata is much smaller than the actual data. However, for short packets, metadata and the actual data are almost of the same size; therefore the usage of conventional methods, such as capacity or ergodic capacity, is highly suboptimal \cite{Durisi_1}. Little work has been done for the short packets domain in the context of ARQ and short packets \cite{Kim2013}. Herein, we show that our proposed algorithms are valid for finite length codewords. This occurs because when power levels are high enough, the dispersion term present in the maximum achievable rate looses dominance. Moreover, as shown in \cite{Durisi}, whenever there is any type of CSI \footnote{Herein we assume CSI at the Receiver side, i.e. CSIR which is a common assumption in URLLC literature and can ve obtained via channel estimation \cite{Durisi_1}.}, the gap to Shannon capacity that is observed due to finite blocklength is closed. 
	
	Further, in \cite{Devassy2014} the authors analyze the performance of ARQ protocol over the fading channel under very simplistic assumptions such as infinite number of transmissions, full buffer capacity,  instantaneous and error free feedback. However, therein the authors do not investigate the impact of power allocation between different ARQ rounds. In \cite{Makki2015}, the authors develop a power allocation scheme for type-I ARQ protocol that minimizes the outage probability only for the case of two transmissions. However, in their scheme they do not guarantee a minimal outage probability level which would be essential in the case of URC.
	
	\subsection{Contributions}
	In this paper, we focus on Machine Type Communications (MTC) that have very stringent requirements on latency and reliability. We show that achieving ultra low latencies under the one shot transmission setup would not be feasible, due to very large power consumption. Furthermore, we show that in medium and high SNR regime, the asymptotic approximation of the outage probability provides a good benchmark for further analysis. 
	
	To mitigate the large power expenditures associated with the one shot transmission, we suggest the implementation of retransmission schemes. We assume that it is possible to evaluate the delay associated with one transmission round and select the maximum number of transmissions so that the delay requirement of the system is satisfied. Specifically, we focus on the problem of optimal power allocation for repetitive and parallel IR-HARQ transmission schemes under block fading channel conditions. For this purpose, we cast an optimization problem to minimize the average power expenditure needed to meet a certain target outage probability. Herein, we extend our initial analysis presented in \cite{me, me1}\footnote{Notice that part of this work was initially presented in IEEE ICC'17 \cite{me} and EUCNC'17 \cite{me1}.} for ARQ and CC-HARQ protocols, to more complex schemes such as IR-HARQ. For this scheme, we first develop a closed form approximation of the outage probability. Next, by proving the convexity of our problem, we show that the obtained solutions are globally optimal.
	
	The main contributions of the paper are as follows
	\begin{itemize}
		\item We develop power allocation algorithms of low (linear) complexity, which would allow us to achieve any target outage probability for repetition and parallel (IR-HARQ) retransmission protocols that would enable operation in the URR and the finite block-length regime at any number of transmissions.
		\item We show that the optimization problem is convex, and that the globally optimal power allocation strategy suggests a transmission with increasing power in each transmission round when ultra low outage probability values are required. 
		\item We provide a closed-form approximation for the outage probability of the IR-HARQ protocol.
		\item We compare the power and efficiency of the algorithms that we have developed and show that the best power saving is achieved by IR-HARQ.
	\end{itemize}
	
	\subsection{Outline}
	The remainder of this paper is organized as follows. In	Section \ref{sc:2}, we introduce the system model and analyze the one shot transmission in the finite block-length regime. Section \ref{4} presents the optimal power allocation algorithms for both repetitive and IR-HARQ retransmission schemes. In Section \ref{6}, we provide comparisons for the performance of the proposed algorithms through some illustrative numerical results. Finally, the main conclusions are summarized in Section \ref{5}.

	\section{Maximum coding rate in finite blocklength} 
	\label{sc:2}
	
	\subsection{System Model}
	\label{model}
	Assume a transmitter-receiver pair communicating under block-fading channel. As in \cite{Yanga}, at the transmitter side, $b_1, b_2, \ldots, b_K$ nats\footnote{To standardize the notation, hereafter we assume that all information is encoded in nats instead of bits. Therefore, all $\log$ is the natural logarithm.
	} are encoded in $c_1, c_2, \ldots, c_{n_c}$. Next, these encoded nats are interleaved and mapped to a constellation $\mathcal{X}$. This results in the stream of modulated symbols $x_1, x_2, \ldots, x_n$. For simplicity, we assume that we map one modulated symbol per channel use. Here, $K$ and $n$ denote the number of information nats and the number of channel uses, respectively. This results in the creation of the packets that will be transmitted. The receiver then fetches the packets and tries to recover the information.
	
	While communicating, the pair can either utilize an open loop setup, or a retransmission protocol. In the second case, the maximum number of transmissions is set to $M$. Whenever the transmitter fetches a negative acknowledgment (NACK) packet, it retransmits based on the protocol that is implemented. It will stop sending packets if it receives an ACK message from the receiver, or if the maximum number of allowed transmissions has been completed.
	
	We consider quasi-static fading channel conditions, in which the channel gain $h$ remains constant for the duration of one packet transmission and changes independently between all the transmission rounds. The motivation is that URLLC and MTC devices usually communicate on short packets where the channel fading coefficient is almost constant through the packet duration \cite{TVT_Shehab}. This assumption simplifies the analysis and serves the goal of obtaining benchmark power allocation for the problem addressed. Moreover, it has been widely accommodated in the literature such as in \cite{Durisi_1}. We analyze the case when the channel coefficient is Rayleigh distributed and $h\sim \mathcal{CN} (0,1)$. Thus the squared envelope of the channel gain is exponentially distributed with mean one. For simplicity we denote  $f_{|h|^2} (z) = e^{-z}$. We assume that the receiver has channel state information while the transmitter knows only the distribution of the channel gains and the information it obtains from the feedback. Then the received signal at the $m^{th}$ round can be written as
	\begin{align}
		\label{eq:1}
		\mathbf{y_m} = \sqrt{\rho_m}h_m\mathbf{x_{m}} + \mathbf{w_{m}},
	\end{align}
	where $\mathbf{x_{m}}$ is the transmitted signal and $\mathbf{w_{m}}$ is the AWGN noise term with noise power $N_0=1$. The term $\rho_m$ is the packet transmitted power. Since the variance of the noise is set to unity, it corresponds to the signal-to-noise ratio (SNR).

	\subsection{One shot transmission}
	\label{3}
	In this section, we briefly summarize the recent results in the characterization of the maximum channel coding rate and outage probability in the finite block-length regime. Further, we evaluate the case of the open loop setup. 
	
	For notational convenience we need to define an $(n, K, \rho, \epsilon)$ code as a collection of
	\begin{itemize}
		\item An encoder $\mathcal{F} : \{1, \ldots, K\} \mapsto \mathcal{C}^n $ which maps the message $ k \in \{1, \ldots, K\}$  into an $n$-length codeword $c_i \in \{c_1, \ldots , c_n\}$ such that the following power constraint ($\rho$) is satisfied:
		\begin{align}
			\label{eq:2}
			\frac{1}{n} \|c_i\|^2 \leq \rho, \forall i.
		\end{align}
		\item A decoder $\mathcal{G} : \mathcal{C}^n \mapsto \{1, \ldots, K\}$ that satisfies the maximum error probability ($\epsilon$) constraint:
		\begin{align}
			\label{eq:3}
			\max_{\forall i} \mathrm{Pr}\left[\mathcal{G}(y) \neq I|I=i\right] \leq \epsilon,
		\end{align}
		where $y$ is the channel output induced by the transmitted codeword according to (\ref{eq:1}).
	\end{itemize}
	The maximum achievable rate of the code is defined as \cite{Yanga}
	\begin{align}
		\label{eq:4}
		R_{max}^*(n, \rho, \epsilon)=\sup \Big\{\frac{\log  K}{n}: \exists(n, K, \rho, \epsilon)~\mathrm{code}\Big\}.
	\end{align}

	For the AWGN channel non-asymptotic lower and upper bounds on the maximum achievable rate have been derived in \cite{Polyanskiy2010}. Recently, a tight approximation for $R_{max}^*(n, \rho, \epsilon)$ has been proposed for sufficiently large number of channel uses (\textit{i.e.} $n>100$) in the case of the quasi-static fading channel \cite{Yanga} and is given by 
	\begin{align}
		\label{eq:5}
		R_{max}^*(n, \rho, \epsilon)\approx C_\epsilon + \operatorname{O}\left( \frac{\log n}{n}  \right),
	\end{align}
	where $C_\epsilon$ is the outage capacity:
	\begin{align}
		\label{eq:6}
		C_\epsilon= \sup \{R: \mathrm{Pr}[\log (1+\rho \cdot z)<R]<\epsilon\}. 
	\end{align}
	Then, by a channel coding rate of $R=\frac{K}{n}$ nats per channel use (ncpu), where $K$ is the information payload, the outage probability is approximated as \cite{Yangk}
	\begin{align}
		\label{eq:7}
		\epsilon (n, R, \rho) &\approx \operatorname{E}_z\left[\operatorname{Q}\left(\frac{C(\rho \cdot z)-\frac{K}{n}}{\sqrt{V(\rho \cdot z)}}\right)\right] \\ &\approx \int_{0}^{\infty} e^{- z  } \operatorname{Q}\left(\frac{C(\rho \cdot z  )-\frac{K}{n}}{\sqrt{V(\rho \cdot z  )}}\right)\mathrm{d}z,
	\end{align}
	where $\operatorname{E}[\cdot]$ denotes the expectation over the channel gain $z $, $\operatorname{Q}(\cdot)$ denotes the Gaussian Q-function, $C(x) = \log(1+x)$ denotes the channel capacity and the channel dispersion is computed as $V(x) = 1 - \tfrac{1}{(1+x)^2}$. However the integral in (8) does not have a closed form solution. Thus, we resort to an approximated closed-form expression as in \cite{Makki2014}
	\begin{align}
		\label{eq:11}
		\epsilon (n, R, \rho) = 1-\frac{\delta}{\sqrt{2\pi}}e^{-\kappa}\left(e^{\sqrt{\frac{\pi}{2\delta^2}}}-e^{-\sqrt{\frac{\pi}{2\delta^2}}}\right),
	\end{align}
	where $\kappa=\frac{e^R-1}{\rho}$ and $\delta=\sqrt{\frac{n\rho^2}{e^{2R}-1}}$. Note that \eqref{eq:11} characterizes the outage probability of a single ARQ round.
	
\begin{figure}[!htb] 
		\centering 
		\includegraphics[trim=0.8cm 0.1cm 1.3cm 0.3cm, clip=true,width=0.8\linewidth]{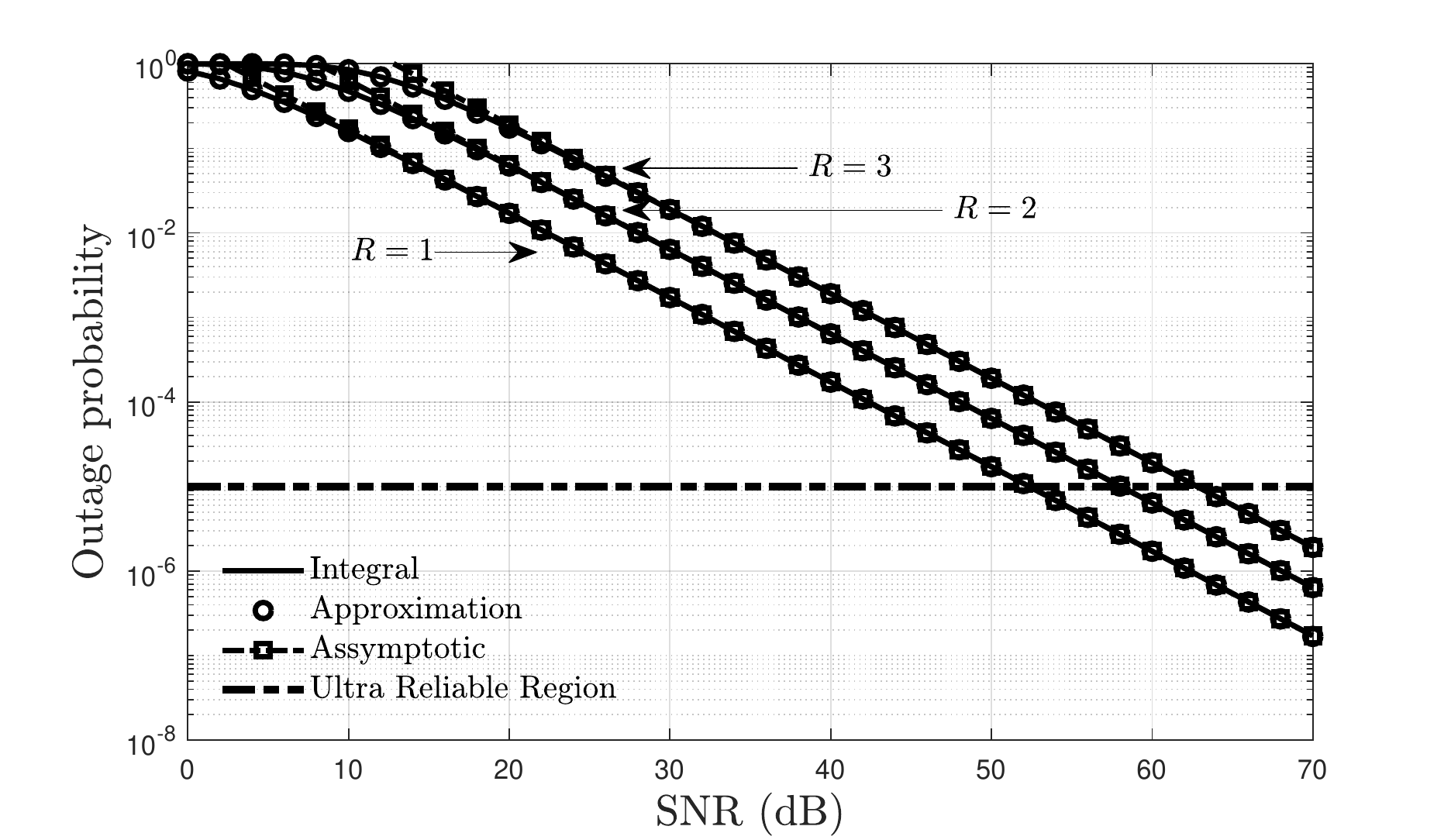}
		\vspace{-4mm}
		\caption{Outage probability for the open loop setup for different channel coding rates.}
		\label{fig:open_loop_setup}
		\vspace{-5mm}
	\end{figure}

	Fig. \ref{fig:open_loop_setup} illustrates the outage probability for the open loop setup, where the message is conveyed in a single transmission, for different channel coding rates. We have fixed the number of channel uses $n=200$ and analyzed the case of mapping $K \in \{200,400,600\}$ information nats. This results in the channel coding rates $R=1$, $R=2$ and $R=3$ ncpu, respectively. We can see that the integral form in (8) and the closed-form approximation in \eqref{eq:11} match well for all the coding rates.
	
	Furthermore, in Fig. \ref{fig:open_loop_setup} we also illustrate the performance of the asymptotic approximation (which we derive next and is given by \eqref{eq:14}). The results show that for $\epsilon < 10^{-2}$ and in the ultra reliable region (URR), where very low outages are required (i.e., $\epsilon < 10^{-5}$), the asymptotic approximation can be used. This is due to the fact that in high SNR the maximum achievable rate (\ref{eq:5}) converges to the one with asymptotically long codewords $R^* (n,\epsilon)=C_{\epsilon}$, where $C_{\epsilon} $ is defined in (\ref{eq:6}). However, the amount of power required to achieve ultra reliability for this setup is quite high (> 50 dB). Herein, when $\rho \rightarrow \infty$ the outage probability in the $m^{th}$ round can be calculated as:
	\begin{align}
		\label{eq:13}
		\epsilon_m=\mathrm{Pr}[\log (1+\rho  z  )<R]=1-e^{-\frac{e^{R}-1}{\rho_m}}.
	\end{align}
	The equality in (\ref{eq:13}) holds for Rayleigh fading channels. Furthermore, by using the first order of Taylor expansion $e^{-x} \approx 1-x$ we can express the asymptotic approximation for the outage probability of the $m^{th}$ round as:
	\begin{align}
		\label{eq:14}
		\epsilon_m=\frac{\phi}{\rho_m},
	\end{align}
	where 
	\begin{align}
		\label{phi}
		\phi=e^{R}-1. 
	\end{align}
	
	For ultra reliable communications, we require to have an outage probability $\epsilon$ very low, while spending as little power as possible. However, Fig. \ref{fig:open_loop_setup} shows that such low outage values are highly unlikely to be obtained when using an open loop setup. Thus, we investigate the possibility of utilizing retransmission mechanisms with optimal power allocation,  in order to obtain an outage probability in the ultra reliable region. 
	
	\section {Optimal Power allocation}
	\label{4}
	In this section we evaluate the impact of repetitive and IR-HARQ retransmission schemes in the outage probability. Specifically, we focus on the analysis of ARQ, CC-HARQ and IR-HARQ protocols. We propose an optimal power allocation scheme that allows us to reach any target outage probability, assuming that we can have up to $M$-transmissions for each of the protocols. 
	
	The problem of interest is to achieve a target outage probability while spending as little power as possible for conveying the information from the transmitter to the receiver. Since we will have multiple transmissions, one approach would be to allocate equal power in each round. This implies that given a certain power budget $\rho_{budget}$, the transmit power in the $m^{th}$ round would be $\rho_m=\frac{\rho_{budget}}{M}$. However, such simplistic approach is shown to be highly inefficient for very low outage probability values \cite{Chaitanya2016, Makki2015}. Thus, we propose a power allocation algorithm in order to minimize the average transmit power of the transmitter which allocates different power levels in each retransmission round. Bearing this in mind, the average transmitted power can be defined as
	\begin{align}
		\label{eq:15}
		\rho_{avg}=\sum_{m=1}^M \rho_m E_{m-1}\mathrm{,}
	\end{align}
	where $M$ is the maximum number of retransmission rounds, $\rho_m$ is the power transmitted in the $m^{th}$ round and $E_{m-1}$ is the outage probability up to the $m-1$ round. Next, we calculate the probability that the packet can not be decoded correctly even after the maximum allowed number of retransmissions. We refer to this as packet drop probability (pdp), and it corresponds to the outage probability up to the $M^{th}$ round ($E_M$). Since we have assumed that all the transmissions of the packets experience independent fading conditions, we can express $E_M$ as
	\begin{align}
		\label{eq:16}
		E_M=\prod_{m=1}^M \epsilon_m \mathrm{,}
	\end{align}
	where $\epsilon_m$ is the outage probability of the of the $m^{th}$ ARQ round
	and can be computed by \eqref{eq:11}, or asymptotically via \eqref{eq:14}. The outage probability before the first transmission, $\epsilon_0=1$. 
	
	Based on our system model introduced in Section \ref{model}, we now formulate the main problem of this paper as follows:
	\begin{Problem}
		The optimal power allocation strategy for repetitive and IR-HARQ retransmission schemes is the solution of
		\begin{equation}
			\begin{aligned}
				\label {eq:17}
				 {\min_{\rho_m}} \ \ \
				 & \mathrm{\rho_{avg}} \\
				 \text{s.t} \ \ \
				 & 0 \leq \rho_{m}, \; 1 \leq m \leq M \\
				& E_M \leq \epsilon
			\end{aligned}
		\end{equation}
		where $\epsilon$ is any target outage probability. 
	\end{Problem}
	
	To obtain a globally optimal analytical solution for this problem we first verify its convexity, since it is a necessary and sufficient condition to utilize Karush-Kuhn-Tucker (KKT) conditions \cite{Boyd-Vandenberghe-04}.
	To do so, we propose the following lemma.  
	
	\begin{lemma}
		\label{lm:1}
		The optimization problem \eqref{eq:17} is convex for the protocols that will be analyzed throughout this paper.
		\begin{proof}
			Please see Appendix \ref{A}.
		\end{proof}
	\end{lemma}
	
	Now, we can write the Karush-Kuhn-Tucker (KKT) conditions to obtain the optimal power allocation strategy for the convex problem (\ref{eq:17}). First, we write the Lagrangian function as 
	\begin{align}
		\label{eq:x}
		\mathcal {L} (\rho_m, \mu_m,\lambda) \!=\! \sum_{m=1}^M \rho_m E_{m-1} + \sum_{m=1}^M \mu_m \! \rho_m+ \lambda(E_M-\epsilon),
	\end{align}
	where $\mu_m$ for $m=1,\ldots,M$  and $\lambda$ are the Lagrangian multipliers. Furthermore, we express the KKT conditions as follows: 
	\begin{enumerate}[label=\textbf{C\theenumi},itemsep=2pt,parsep=2pt,topsep=2pt,partopsep=2pt]
		\item $\frac{\partial{\mathcal{L}}}{\partial {\rho_m}}=0,~ m=1,\ldots,M$,
		\label{c1}	
		\item $\mu_m \geq 0,~ m=1,\ldots,M$,
		\label{c2}
		\item $\mu_m\rho_m=0,~ m=1,\ldots,M$,
		\label{c3}
		\item $E_M-\epsilon=0$.
		\label{c4}
	\end{enumerate}
	Note that the target is to minimize the transmit power. Thus, it is straight forward to infer that the reliability constraint is optimaly achieved at equality since more power is needed to achieve lower error and higher reliability. Hereafter, we begin the solution of \eqref{eq:17} for the repetition and IR-HARQ retransmission schemes.
	
	\subsection{Repetitive retransmission schemes}
	\label{rep}
	In repetitive retransmission schemes the transmitter sends the same information nats in each round. Here, we present the power allocation algorithms for two retransmission protocols that utilize repetitive schemes, respectively ARQ and CC-HARQ. Furthermore, we discuss about the main differences between them. 
	
	\subsubsection{ARQ}
	\label{bbb} 
\begin{figure}[!b] 
		\vspace{-10mm}
		\centering
		\includegraphics[trim=0cm .5cm 0cm 0cm, clip=true,width=0.8\linewidth]{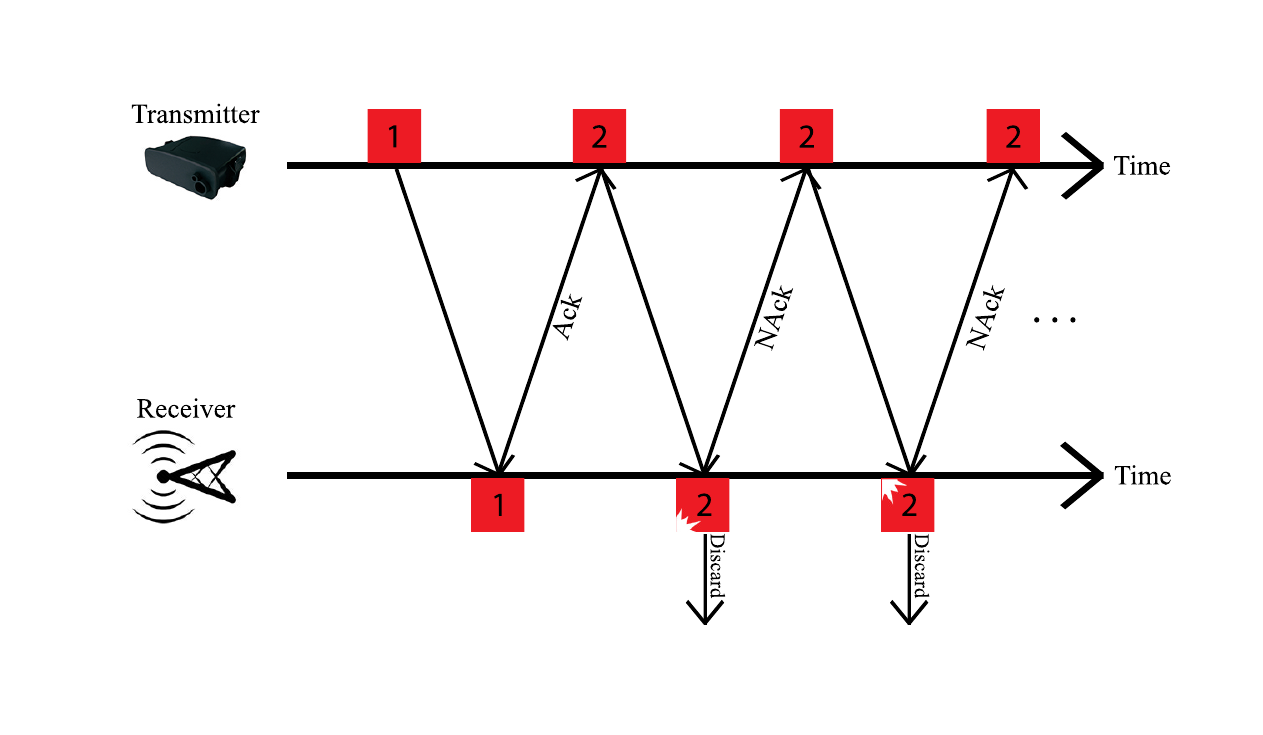} \vspace{-5mm}
		\caption{The setup for ARQ protocol. The transmitter sends the first packet and waits for the ACK from the receiver. Then it sends packet $2$. If the receiver can not decode the packet, it discards the packet and asks for retransmission.}
		\label{fig:ARQ} \vspace{-4mm}
	\end{figure}

	Its principle is shown in Figure. \ref{fig:ARQ}. 
	The transmitter sends the whole packet in each transmission round and stops when the maximum number of transmissions $M$ has been achieved, or when it receives confirmation that the packet has been successfully decoded. The receiver makes the decisions only based on the last packet he has received, and discards the earlier packets. 
	
	For this protocol, the optimal power terms can be obtained recursively backward using the lagrangian $\lambda$ as described in Theorem \ref{th1}. 
	\begin{theorem}
		\label{th1}
		The optimal power terms for the ARQ protocol are
		\begin{align}
			\label{eq:21}
			\rho_M&=\sqrt{ M\lambda \phi},\\
			\label{eq:28}
			\rho_m&=\sqrt{2\phi\rho_{m+1}}, \; 1\leq m<M,
		\end{align}
		where $\phi$ and $\lambda$ are given from \eqref{phi} and \eqref{eq:39} respectively. 
	\end{theorem}
	
	\begin{proof}
		The outage probability for the $m^{th}$ ARQ round and can be computed by \eqref{eq:11}, or asymptotically via \eqref{eq:14}. We start the solution for problem \eqref{eq:17} by writing the Lagrangian function, which is computed as
		\begin{align}
			\label{eq:18}
			\mathcal {L} (\rho_m, \mu_m,\lambda) \!=\! \!\sum_{m=1}^M\! \rho_m E_{m-1}^{ARQ}\! + \!\sum_{m=1}^M \! \mu_m \rho_m\!+ \!\lambda(E_M^{ARQ}\!-\!\epsilon).
		\end{align}
		
		Next we analyze the KKT conditions. From \ref{c1}, we write the derivative of the Lagrangian function $\mathcal {L} (\rho_m, \mu_m,\lambda)$ with respect to the power $\rho_m$ as
		\begin{align}
			\frac{\partial \mathcal {L} (\rho_m, \mu_m,\lambda)}{\partial \rho_m }=&\left(\frac{\phi^{m-1}}{\prod_{i=1}^{m-1}\rho_i} - \sum_{i=1}^{M-m}\frac{\rho_{m+i}\phi^{m+i-1}}{\rho_m^2\prod_{j=1, j \neq m}^{m+i-1}\rho_j}\right) - \mu_m - \lambda\left(\frac{\phi^M}{\rho_m^2 \prod_{i=1, i \neq m}^{M}\rho_i}\right).
			\label{eq:19}
		\end{align}
		We relax temporarily the non-negative condition on the transmitted power terms $\rho_m$. This implies that, $\mu_m = 0$ for $m =1,...,M$. Now, using \eqref{eq:19} and $\mu_M = 0$, we can write \ref{c1} for $m = M$ as 
		\begin{align}
			\label{eq:20}
			\frac{\partial \mathcal {L} (\rho_m, \mu_m,\lambda)}{\partial \rho_M }\!=\!\frac{\phi^{M-1}}{\prod_{i=1}^{M-1}\rho_i} \!-\! \lambda\left(\frac{\phi^M}{\rho_M^2\prod_{i=1,i\neq m}^M \rho_m}\right)\!\!=\!0.
		\end{align}
		After some algebraic manipulations of \eqref{eq:20} we obtain the transmit power at the $M^{th}$ ARQ round as in \eqref{eq:21}. Similarly, by substituting $m=M-1$ in \eqref{eq:19} and using $\mu_{M-1} = 0$, we can rewrite \ref{c1} for $m= M-1$ as
		\begin{align}
			\label{eq:23}
			\frac{\partial \mathcal {L} (\rho_m, \mu_m,\lambda)}{\partial \rho_{M-1} }&=\left(\frac{\phi^{M-2}}{\prod_{i=1}^{M-2}\rho_i}-\frac{\rho_M\phi^{M-1}}{\rho_{M-1}^2\prod_{i=1}^{M-2}\rho_i}\right) - \lambda\left(\frac{\phi^M}{\rho_{M-1}^2\prod_{i=1,i\neq M-1}^M \rho_i}\right)=0,
		\end{align}
		which can be simplified, yielding
		\begin{align}
			\label{eq:24}
			\rho_{M-1}=\sqrt{\rho_M\phi+\frac{M\lambda\phi^2}{\rho_M}}.
		\end{align}
		Following a similar approach, we obtain the following relationship for the case of $m = M-2$ 	
		\begin{align}
			\label{eq:25}
			\rho_{M-2}=\sqrt{\rho_{M-1}\phi+\frac{\rho_M\phi^2}{\rho_{M-1}}+\frac{M\lambda\phi^3}{\rho_{M-1}\rho_M}}.
		\end{align}
		We can continue this procedure for all $m \in \{1, \ldots, M\}$ and the results can be summarized as:
		\begin{align}
			\rho_M&=f(\lambda),\\
			\label{b}
			\rho_{M-1}&=f(\lambda, \rho_M), \\
			\label{a}
			\rho_{M-2}&=f(\lambda, \rho_M, \rho_{M-1}), \\
			\vdots \nonumber\\
			\rho_{1}&=f(\lambda, \rho_M, \ldots, \rho_3, \rho_2). 
		\end{align}
		
		At this point, we can easily verify now that the obtained power values are positive and $\rho_m$ cannot be further
		minimized. By utilizing a method that is similar to the backward substitution approach \cite[App. C.2]{Boyd-Vandenberghe-04}, we can obtain a relationship between the power terms $\rho_m$ as follows: {\it i)} by substituting $M \lambda \phi = \rho_M^2$ (see (\ref{eq:21})) in \eqref{eq:24} (or equivalently in \eqref{b}) we evaluate $\rho_{M-1}$ as $\rho_{M-1} = \sqrt{2\phi \rho_M}$. {\it ii)} $\rho_{M-2}$ is evaluated by substituting  $\sqrt{M\lambda \phi} = \rho_M$ and  $\sqrt{2\phi \rho_M} = \rho_{M-1}$ in \eqref{a}. {\it iii)} By continuing this procedure we can express the optimal transmit power in the $m^{th}$ round as in \eqref{eq:28}.
		
		Based on \eqref{eq:28}, we can easily verify now that the obtained power values $\rho_m$ are all positive. Further, since $\rho_M$ is a function of $\lambda$ (see \eqref{eq:21}) and using \eqref{eq:28}, it is clear that each $\rho_m$ is a function of $\lambda$. Thus, all that remains is to compute the Lagrangian multiplier $\lambda$. For this purpose, we utilize the outage constraint in (\ref{eq:17}) (\ref{c4}). First, we substitute $\rho_m$ for $m =1,\ldots,M$ in \eqref{eq:16} to obtain $E_M^{ARQ}$ as
		\begin{align}
			\label{eq:29}
			E_m^{ARQ}=\frac{\phi^M}{\prod_{m=1}^M \rho_m}=\epsilon,
		\end{align}
		where $\rho_m$ is given by 
		\begin{align}
			\label{eq:32}
			\rho_m=\sqrt{2^{a(m)}  \phi^{b(m)} (M \lambda)^{c(m)}}.
		\end{align}
		In \eqref{eq:32}, we can compute the exponents 
		\begin{align}
			\label{a(m)}
			a (m)&={2-2^{-(M-m-1)}}\\
			\label{b(m)}
			b(m)&={2-2^{-(M-m)}}\\
			\label{c(m)}
			c(m)&={{2^{-(M-m)}}}
		\end{align}
		Finally, we compute $\lambda$ by equating $E_M^{ARQ}$ to the outage target $\epsilon$ based on \ref{c4}:
		
		\begin{align}
			\label{eq:39}
			\lambda=\left(\frac{\phi^{o(M)}}{M^{o(M)} \, \epsilon^{p(M)}\,2^{q(M)}}\right)^{\tfrac{1}{o(M)}},
		\end{align}
		where
		\begin{align}
			\label{o(M)}
			o(m) &= 2^m-1,\\
			\label{p(M)}
			p(m) &= 2^m,\\
			\label{q(M)}
			q(m) &= [(m-2)2^m+2].
		\end{align}
	\end{proof}
	
	Finally, we utilize the results presented in Theorem \ref{th1} and propose the following power allocation algorithm for ARQ protocol.

	\begin{algorithm}
		\caption{Power allocation for ARQ}\label{Type-I ARQ}
		\begin{algorithmic}[1]
			\item {\bf Inputs:} $\phi$, $M$.
			\item {\bf Compute:} $a(m)$ as in \eqref{a(m)}.
			\item {\bf Compute:} $b(m)$ as in \eqref{b(m)}.
			\item {\bf Compute:} $c(m)$ as in \eqref{c(m)}.
			\item {\bf Compute:} $o(M)$ as in \eqref{o(M)}.
			\item {\bf Compute:} $p(M)$ as in \eqref{p(M)}.
			\item {\bf Compute:} $q(M)$ as in \eqref{q(M)}.
			\item {\bf Compute:} $\lambda$ as in \eqref{eq:39}.
			\item {\bf Compute:} $\rho_M$ as in \eqref{eq:21}.
			\item {\bf while} all power terms are not found {\bf do}
			\item \quad{\bf Compute:}  $\rho_m$ as in \eqref{eq:28}.
			\item \quad{\bf Decrease:} $m$ by 1.
			\item {\bf end while}.
			\item {\bf Outputs:} All the power terms $\rho_m$.
		\end{algorithmic}
	\end{algorithm}
	
	From Theorem \ref{th1}, it is obvious that the analytical solution to obtain the optimal power allocation scheme for a large number of retransmissions would become a cumbersome task. Furthermore, in general the ultra reliable systems are delay-limited. For this purpose, in Appendix \ref{F}, we propose a simpler analytical solution which is valid only for the case of $M=2$ transmissions. 
	
	Another important observation we can note from Theorem \ref{th1} is the following:
	\begin{theorem}
		\label{th2}
		In the ultra reliable region, the optimal power allocation strategy suggests a transmission with increasing power in each ARQ round.
	\end{theorem}
	\begin{proof}
		Please refer to Appendix \ref{Appendix B}.
	\end{proof}	
	
	Notice that this result fully matches the intuition. Presuming that the delay requirements are met (by setting $M$ accordingly), our goal is to achieve a target outage probability by spending as little power as possible. For this purpose, we transmit first with low power. If the channel conditions are good then the transmission will be successful, and a large amount of power is saved (as will be pointed out later in Figure \ref{fig:Power_terms_vs_outage}). If it fails, then retransmissions are carried out until an ACK is received, or the maximum allowed number of transmissions is reached. 
	
	\subsubsection{CC-HARQ}
	
\begin{figure}[!htb] 
		\centering
		\includegraphics[trim=0cm .5cm 0cm 0cm, clip=true,width=0.8\linewidth]{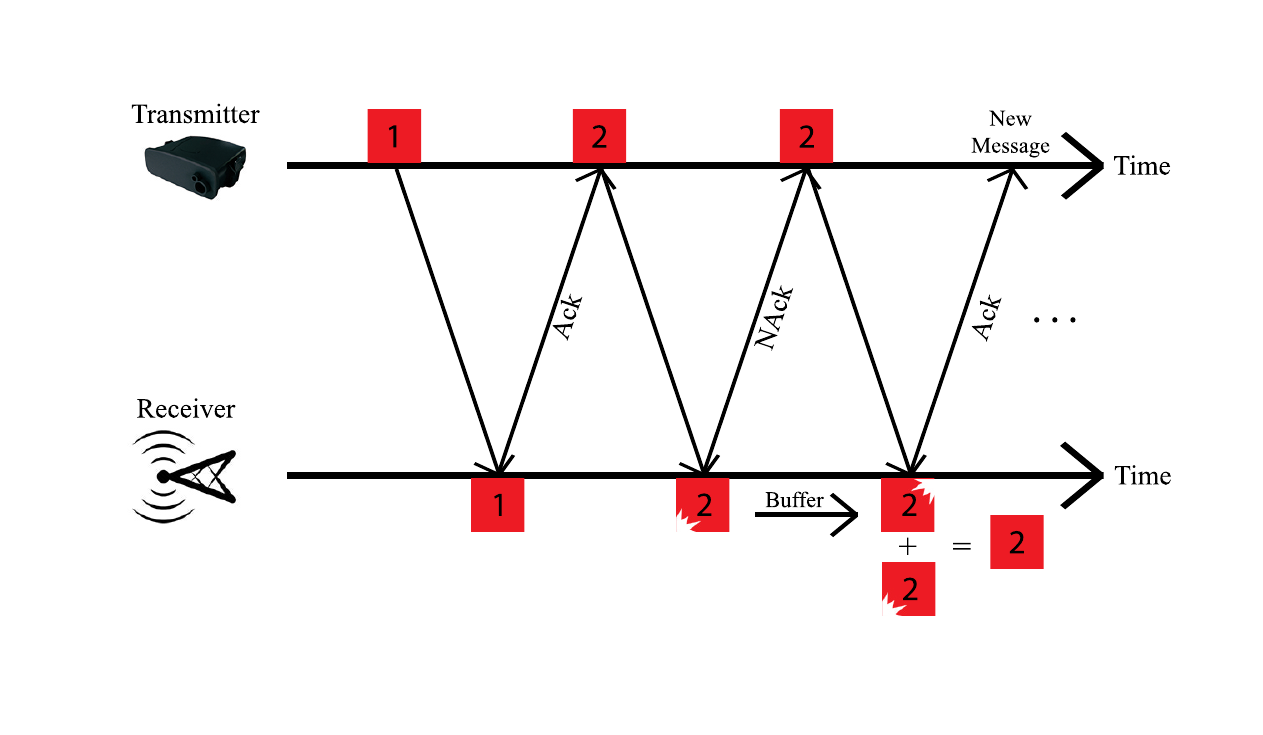} \vspace{-6mm}
		\caption{The setup for CC-HARQ protocol. The transmitter sends first packet $1$ and waits for the ACK from the receiver. Then it sends packet $2$. If the receiver can not decode the packet, it buffers the packet and asks for retransmission. Then, when it receives packet $2$ once more, it combines it with the buffered packet to extract the information.}
		\label{fig:CC-HARQ} \vspace{-4mm}
	\end{figure}

	Another member of the repetitive retransmission schemes is CC-HARQ protocol.  This protocol is widely implemented in several standards (\textit{i.e.} HSDPA, LTE). Its principle is illustrated in Figure \ref{fig:CC-HARQ}. Again, the entire packet is transmitted in each round.  
	However, unlike the previous protocol we analyzed, all the received packets are buffered and the receiver performs MRC to enhance the SNR. When compared to the ARQ protocol, besides the diversity gains, this approach also provides combining gains.
	
	For this protocol, in \cite{Tumula2} the authors have derived a very tight approximation for $E_M$ as
	\begin{align}
		\label{eq:33}
		E_{M}^{CC-HARQ}=\frac{\phi^M}{M! \prod_{m=1}^{M}\rho_m}.
	\end{align}
	Bearing this in mind, we next introduce the optimal power terms for CC-HARQ in the following theorem.
	
	\begin{theorem}
		For the CC-HARQ protocol, we can find the optimal power terms are
		\begin{align}
			\rho_M&=\sqrt{\phi\lambda},\\
			\rho_m&=\sqrt{\frac{2\phi\rho_{m+1}}{m}},
			\label{eq:36}
		\end{align}
		where $\phi$ and $\lambda$ are given from \eqref{phi} and \eqref{eq:39} respectively.
	\end{theorem}
	
	\begin{proof}
		Please refer to Appendix \ref{C}.
	\end{proof}
	
	Finally, for CC-HARQ protocol we propose the following power allocation algorithm.
	
	\begin{algorithm}
		\caption{Power allocation for CC-HARQ}\label{CC-HARQ}
		\begin{algorithmic}[1]
			\item {\bf Inputs:} $\phi$, $M$.
			\item {\bf Compute:} $a(m)$ as in \eqref{a(m)}.
			\item {\bf Compute:} $b(m)$ as in \eqref{b(m)}.
			\item {\bf Compute:} $c(m)$ as in \eqref{c(m)}.
			\item {\bf Compute:} $o(M)$ as in \eqref{o(M)}.
			\item {\bf Compute:} $p(M)$ as in \eqref{p(M)}.
			\item {\bf Compute:} $q(M)$ as in \eqref{q(M)}.
			\item {\bf Compute:} $\lambda$ as in \eqref{eq:40}.
			\item {\bf Compute:} $\rho_M$ as in \eqref{j}.
			\item {\bf while} all power terms are not found {\bf do}
			\item \quad{\bf Compute:}  $\rho_m$ as in \eqref{eq:36}.
			\item \quad{\bf Decrease:} $m$ by 1.
			\item {\bf end while}.
			\item {\bf Outputs:} All the power terms $\rho_m$.
		\end{algorithmic}
	\end{algorithm}
	
	\subsection{Parallel retransmission schemes}
	\label{simple}
	Here, we evaluate the effect of implementing parallel (IR-HARQ) retransmission schemes, wherein the transmitter sends different and jointly designed packets for each message. When compared to the repetitive retranmsission schemes discussed in Section \ref{rep}, besides the diversity and combining gains, this approach provides also coding gains.  
	
\begin{figure}[!htb] 
		\centering \vspace{-5mm}
		\includegraphics[width=0.8\linewidth]{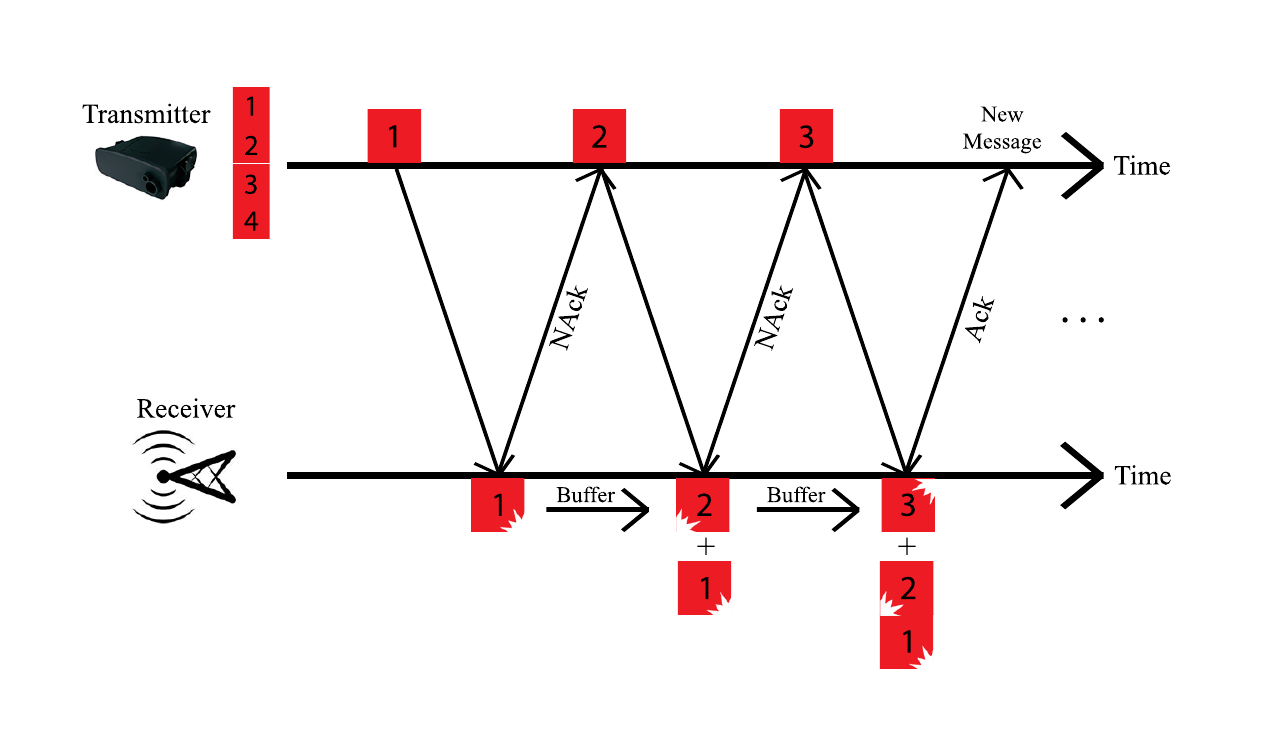} \vspace{-5mm}
		\caption{The setup for IR-HARQ protocol. The transmitter splits the message into several packets. It sends first packet $1$ and waits for reply from the receiver. If it receives NACK packet, then it sends more redundancy through packet $2$ and so on. Meanwhile, the receiver buffers the packets and combines them to extract the information. }
		\label{fig:IR_HARQ} \vspace{-4mm}
	\end{figure}

	The final retransmission protocol that we analyze in this paper is IR-HARQ. Its operating principle is shown in Fig. \ref{fig:IR_HARQ}, where we note that 
	the transmitter splits the parent codeword into $M$ sub-codewords of equal length. Each of these codewords is transmitted during one round. The receiver buffers all the received packets and utilizes them to recover the information. For Gaussian codebook, the mutual information that is gathered would be 
	\begin{align}
		\label{eq:49}
		I= \frac{1}{M} \sum_{m=1}^{M} \log \left(1+\rho_m z_m\right).  
	\end{align}
	After making some manipulations we can compute the outage probability as
	\begin{align}
		\label{eq:50}
		E_M^{IR-HARQ}=\mathrm{Pr} \left[\sum_{m=1}^{M} \log \left(1+\rho_m z_m \right) < MR\right].
	\end{align}
	
	In \cite{Laneman} the author proves a theorem which can be used to approximate \eqref{eq:50}. There, he provides an integral form approximation of the outage probability when equal power is allocated in each retransmission round. Herein, we extend those results and provide closed-form approximation for the outage probability when different power levels are allocated in each IR-HARQ round. It can be computed as given in the following lemma.  
	
	\begin{lemma}
		For the proposed setup, the outage probability of the IR-HARQ protocol can be approximated as
		\begin{align}
			\label{eq:41}
			E_M^{IR-HARQ} \approx \frac{\psi_M}{\prod_{m=1}^M \rho_m}.
		\end{align}
	\end{lemma}
	
	\begin{proof}
		Please refer to Appendix \ref{D}.
	\end{proof}
	
	At this point, all that remains is to obtain the optimal power allocation strategy. For this purpose, we introduce the following theorem.
	
	\begin{theorem}
		\label{th-ir}
		For the IR-HARQ protocol, the optimal power terms are
		\begin{align}
			\rho_M&=\sqrt{\frac{\lambda M \psi_M}{\psi_{M-1}}}\\
			\rho_m&=\sqrt{\frac{2\rho_{m+1}\psi_m}{\psi_{m-1}}}.
			\label{eq:45}
		\end{align}
		\begin{proof}
			Please refer to Appendix \ref{E}.
		\end{proof}
	\end{theorem}
	
	Finally, we can utilize the results in Theorem \ref{th-ir} and formulate the optimal power allocation algorithm.
	
	\begin{algorithm}
		\caption{Power allocation for IR-HARQ}\label{IR-HARQ}
		\begin{algorithmic}[1]
			\item {\bf Inputs:} $R$, $M$.
			\item {\bf while} all values of $\psi_m (R)$ are not found {\bf do}
			\item \quad{\bf Compute:}  $\psi_m$ as in \eqref{eq:98}.
			\item \quad{\bf Increase:} $m$ by 1.
			\item {\bf end while}.
			\item {\bf Compute:} $a(m)$ as in \eqref{a(m)}.
			\item {\bf Compute:} $c(m)$ as in \eqref{c(m)}.
			\item {\bf Compute:} $d(i)=2^{-i}$.
			\item {\bf Compute:} $o(M)$ as in \eqref{o(M)}.
			\item {\bf Compute:} $p(M)$ as in \eqref{p(M)}.
			\item {\bf Compute:} $q(M)$ as in \eqref{q(M)}.
			\item {\bf Compute:} $\lambda$ as in \eqref{eq:48}.
			\item {\bf Compute:} $\rho_M$ as in \eqref{l}.
			\item {\bf while} all power terms are not found {\bf do}
			\item \quad{\bf Compute:}  $\rho_m$ as in \eqref{eq:45}.
			\item \quad{\bf Decrease:} $m$ by 1.
			\item {\bf end while}.
			\item {\bf Outputs:} All the power terms $\rho_m$.
		\end{algorithmic}
	\end{algorithm}

	\section{Numerical analysis}	\label{6}
	
	In this section we provide further results for the performance of the proposed optimal power allocation algorithms. First, we analyze the behavior of the power terms that will be transmitted in each round as a function of the target outage probability. Further, we show that the utilization of the power allocation algorithms provides on average, large gains when compared to the open loop setup. Also, we show that as the maximum number of transmission increases, the average power that is spent decreases. Finally, we evaluate the gains of the proposed power allocation algorithms under the assumption of maximum power expenditure. 
	
\begin{figure} 
		\centering \vspace{-3mm}
		\includegraphics[trim=0cm 0cm 0cm 0cm, clip=true,width=0.83\linewidth]{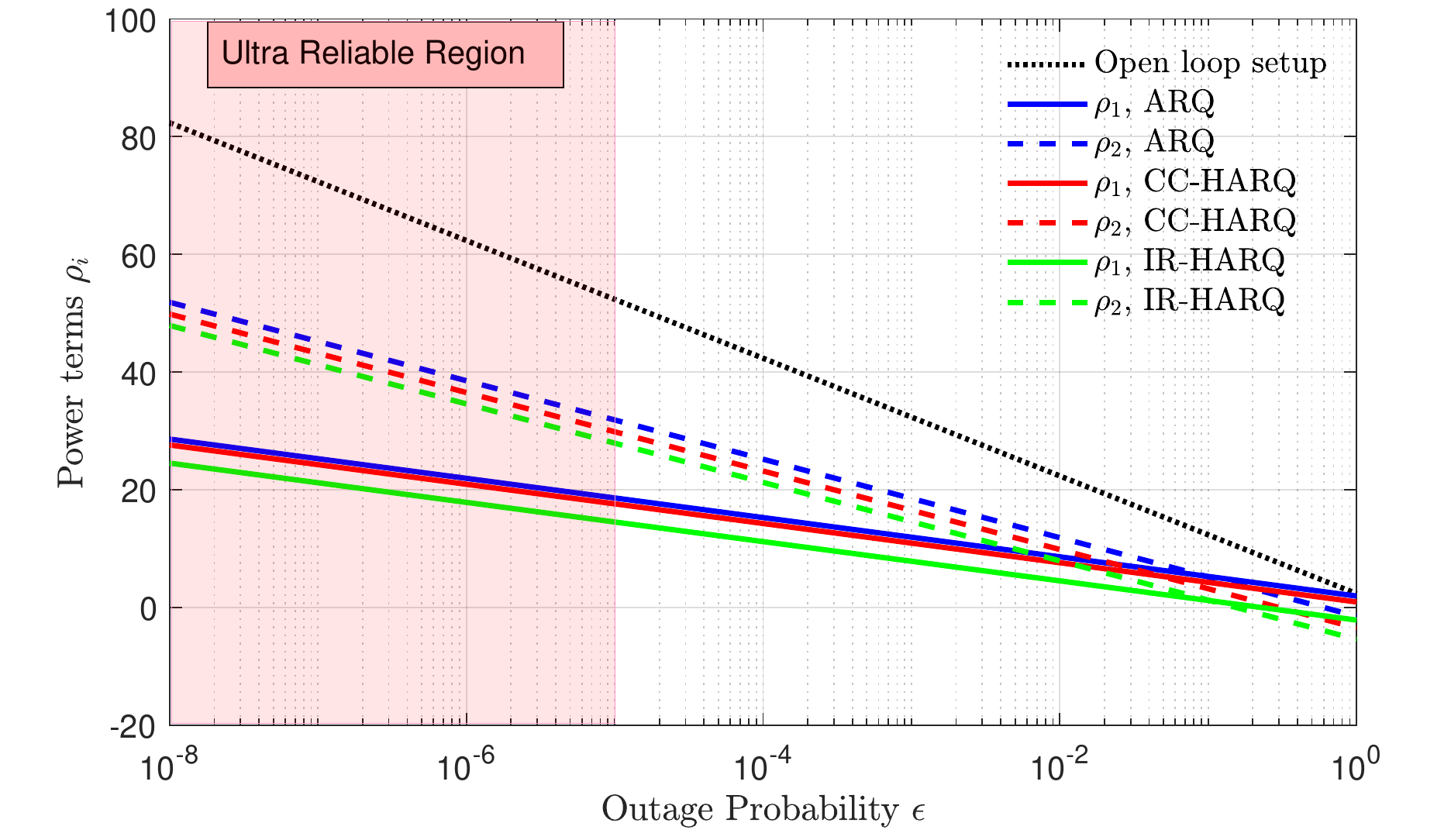} \vspace{-5mm}
		\caption{Transmit power in each round to achieve a target outage probability for rate $R=1$ ncpu.}
		\label{fig:Power_terms_vs_outage} \vspace{-5mm}
	\end{figure}

	In Fig. \ref{fig:Power_terms_vs_outage} we illustrate the variation of transmit power $\rho_m$ in each round versus the outage probability target $\epsilon$ for the case when we have a maximum of two transmissions. The channel coding rate is set to $R=1$ ncpu. The results are obtained by implementing the protocols derived in sections \ref{rep} and \ref{simple}. The first observation we can make is that the IR-HARQ protocol gives the best performance in terms of saving power. Further, we notice that despite the protocol that is implemented both power terms are lower than the open loop transmission which is shown in Fig. \ref{fig:open_loop_setup}. Moreover, if the first round is successful (i.e when the channel conditions are good), then the power gain with respect	to the open loop setup would be very large. Approximately, we save  $30-35$ dB (depending on the protocol) for $\epsilon = 10^{-5}$, which corresponds to the start of the ultra reliable region. Then, the more stringent the reliability requirements are, the more power we save with respect to the open loop setup. We observe that in the URR, the first power term is lower than the second power term. Notice that this result is coherent with what we obtained analytically in Theorem \ref{th2}. 
	
\begin{figure}[!h]
		\centering \vspace{-4mm}
		\includegraphics[trim=0cm 8cm 0cm 7cm, clip=true,width=0.85\linewidth]{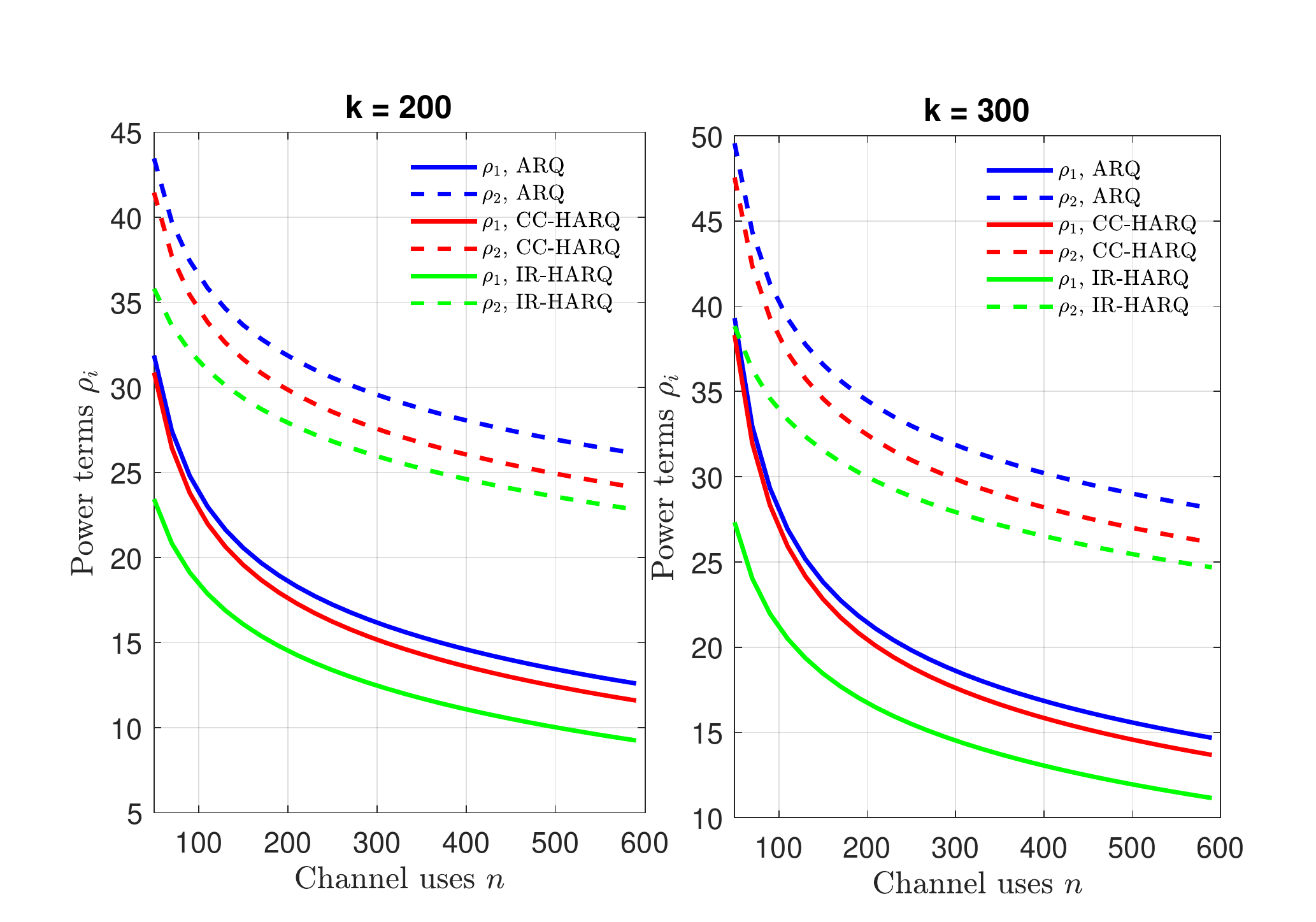} \vspace{-7mm}
		\caption{Power terms as a function of the number of channel uses for different number of information nats.}
		\label{fig:power_terms_vs_n_and_k}
		\vspace{-6mm}
	\end{figure}

	Fig. \ref{fig:power_terms_vs_n_and_k} illustrates the behaviour of the power terms as a function of the number of channel uses when $M=2$. Here we set the number of information nats $K$ to $200$ and $300$ and the target error probability to $10^{-5}$ \footnote{Note that 3GPP defines URLLC requirements for a pyaload od 32 bytes and 99.999\% reliability.}. First we observe that both power terms decrease as we increase the number of channel uses. Secondly we notice that when we increase the coding rate, we have to transmit with higher power in each transmission round. Morever, the figure verifies that IR-HARQ is the most energy efficient scheme as it consumes the least power.
	
\begin{figure} 
		\centering \vspace{-2mm}
		\includegraphics[trim=0cm 0cm 0cm 0cm, clip=true,width=0.72\linewidth]{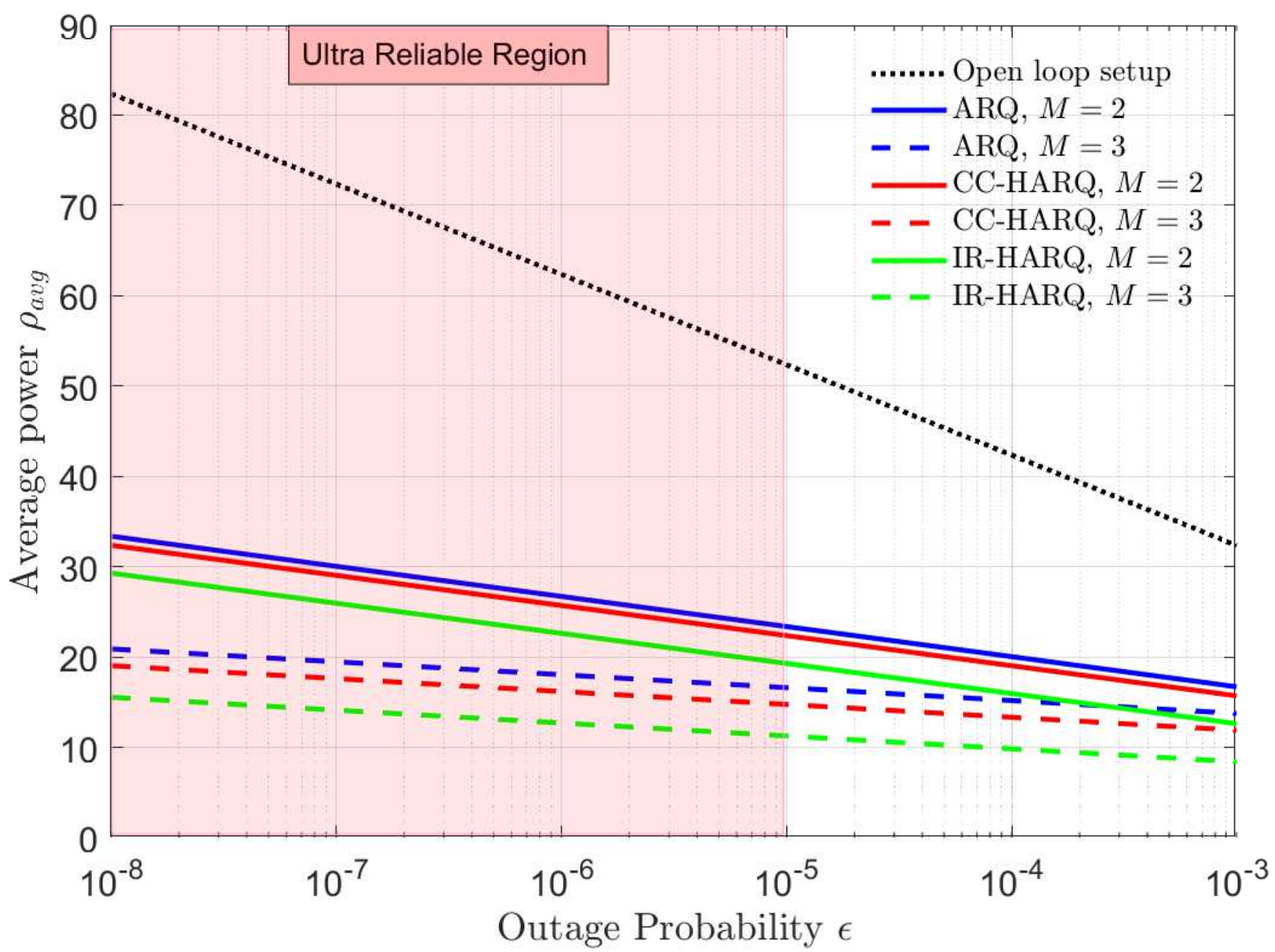} \vspace{-5mm}
		\caption{Average power required to achieve a target outage probability for rate $R=1$ ncpu when the maximum number of transmissions is fixed to $M=2$ and $M=3$.}
		\label{fig:average_power} \vspace{-2mm}
	\end{figure}

	Next, in Fig. \ref{fig:average_power} we evaluate the average power as in \eqref{eq:15} for each proposed scheme and for the scenario when the number of transmissions is increased. To attain this figure, we set $M=2,3$ and $R=1$ ncpu. From it, we notice that the amount of power that is saved per transmission on average is significant, especially when compared to the open loop setup. Further, by comparing the result of Figure \ref{fig:average_power} with the results in Figure \ref{fig:Power_terms_vs_outage}, we notice that as we increase the number of transmissions, we save more power on average. For example, the average power consumption per transmission to achieve an error of $10^{-7}$ using 3 transmissions via IR-HARQ is 10 dB where the total power would be 15 dB. This power consumption level is very low compared to the power hungry 70 dB open loop setup in this case. 
	
\begin{figure} 
		\centering \vspace{-2mm}
		\includegraphics[trim=0cm 0cm 1cm 0cm, clip=true,width=0.83\linewidth]{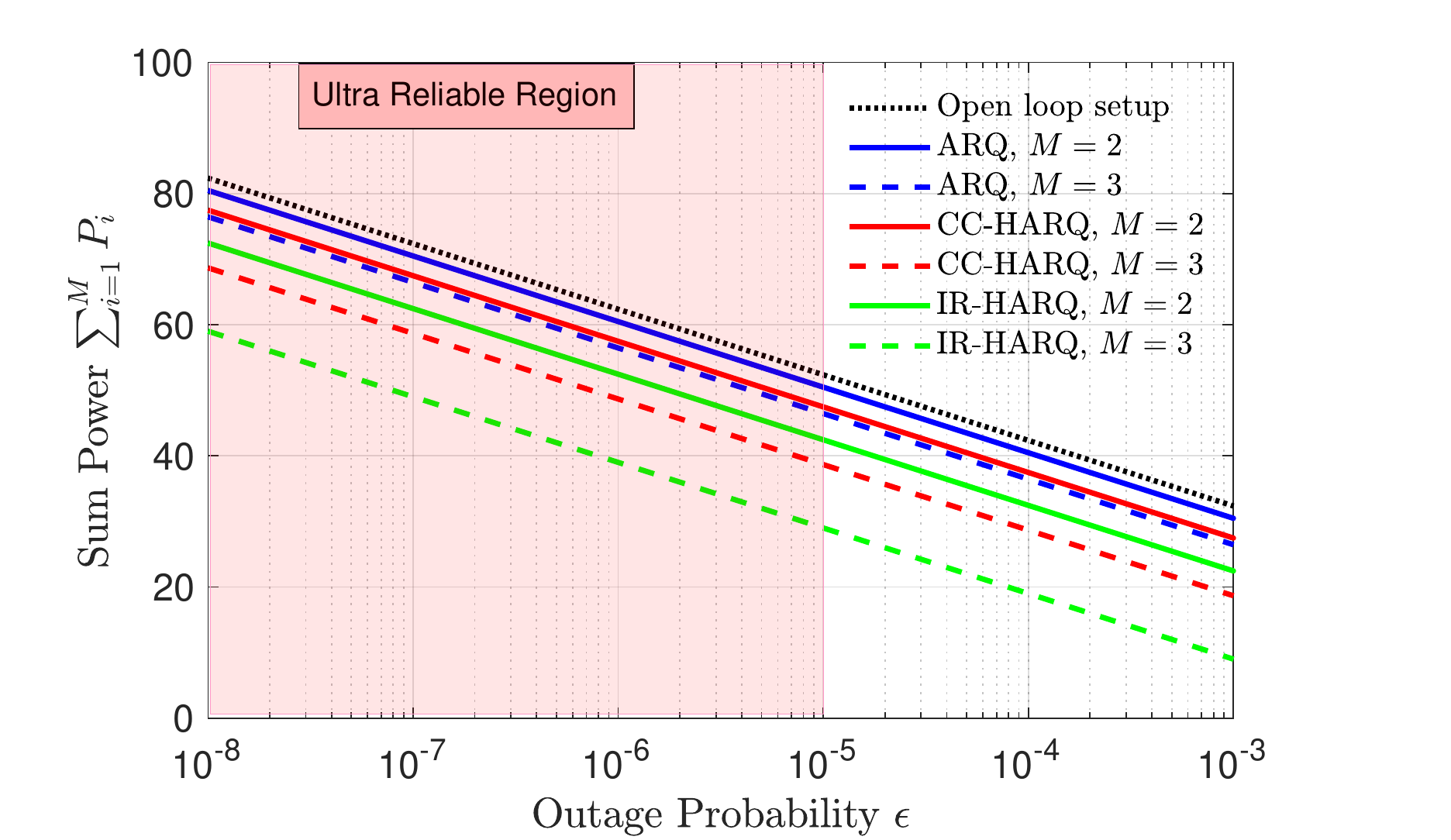} \vspace{-6mm}
		\caption{Maximum power that will be spent to achieve a target outage probability for rate $R=1$ ncpu when the maximum number of transmissions is fixed to $M=2$ and $M=3$.}
		\label{maximum_power_plot} \vspace{-3mm}
	\end{figure}

	Fig. \ref{maximum_power_plot} evaluates the maximum power expenditure of our protocols in the case of $M \in \{2, 3\}$ transmissions and fixed channel coding rate $R=1$ ncpu. To obtain the plot, we assume to have a worst case scenario, where all the transmissions are exhausted. From it, we observe that the proposed algorithms again allow us to save power when compared to the open loop setup. The largest power gains, are again given from the IR-HARQ protocol. Notice that when $M=3$, we can save over $20$ dB by implementing this protocol. Furthermore, we observe that as we increase the number of transmissions, power consumption is reduced. This happens due to the fact that the diversity and combining gains become higher. Also, notice that in the case of ARQ protocol, the power gain when the number of transmissions increases is not as large as the other two protocols. This happens due to the fact that ARQ does not benefit from combining gains. 
	
	\section{Conclusions and future work}
	\label{5}
	In this work, we showed that operation in the URR would not be feasible under the open loop transmission setup, since very large powers are required to achieve ultra-reliability. For this reason, repetitive and parallel retransmission schemes can be implemented. Specifically, we analyzed three popular protocols that embrace these schemes, such as ARQ, CC-HARQ and IR-HARQ. In the case of IR-HARQ, we proposeed a closed form approximation for the outage probability, which was later utilized in our derivations. For all three protocols, we proposed globally optimal power allocation algorithms with low complexity, that guarantee operation anywhere in the ultra reliable region while spending minimal power. We showed that the optimal power allocation strategy to operate in the URR suggests transmission with incremental power in each round. Furthermore, we showed that the best power saving is attained by the IR-HARQ. 
	
	It is obvious that when comparing IR-HARQ and repetitive retransmission schemes, there is the classical trade-off between the system complexity and performance. For example, in applications in which the block-length is very small (e.g. $n \approx 100$), repetitive protocols would be more suitable, since the utilization of IR-HARQ would reduce even more the blocklength which may lead to high outage, and thus larger number of transmissions. Besides, the existing mathematical framework for such short blocklengths does not hold ($n>100$ \cite{Polyanskiy2010}), thus the need for tractable and tighter approximation and bounds for this region. This is further reinforced by the fact that the nodes of such systems generally are very simple sensors. Implementing complex coding and logic in them, would result in a large increase of their cost and would severely affect their battery lifetime, which in these applications corresponds to the device lifetime. However, in other applications the data volumes to be transmitted are far larger, and the packet sizes can be larger (e.g. $n \approx 500$). Generally, the nodes communicating in these systems can afford extra complexity. Therefore, in these type of applications the utilization of IR-HARQ scheme would provide a better performance.
	
	As future work, we intend to analyze the optimal power allocation scheme when there are limitations on the maximum power expenditure. Furthermore, it would be interesting to analyze how the utilization of multi-antenna systems would affect the power expenditure. 
	
	\appendices
	
	\section{Proof of Lemma 1.}
	\label{A}
	A function is convex if both the objective function and the constraint set are convex. Based on \eqref{eq:14}, \eqref{eq:16}, \eqref{eq:33} and \eqref{eq:41} it is straightforward to show that proving the convexity of our problem reduces to proving the convexity of $f(\xi_1, \xi_2, \ldots, \xi_M)=\frac{1}{\xi_1 \xi_2 \ldots \xi_M}$. Notice that $f$ can be written as a composition of two functions
	\begin{align}
		h(y) &= \frac{1}{y},\\
		g(\xi_1, \xi_2, \ldots \xi_M) &= \xi_1 \xi_2 \ldots \xi_M, \text{where $\xi \in R^M$.}
	\end{align}
	
	The function $g$ is concave in $R^M$ and $h$ is convex decreasing function for $y \in R^+$. Next, we analyze the extension value extendibility $\tilde{h}$ of $h$. For this purpose, since $h$ is a convex function, we assign the value $\infty$ to all the points that are not in the domain of $h$. Note that
	\begin{align}
		\label{key}
		\lim\limits_{z \rightarrow 0^-} h =\infty \geq \lim\limits_{z \rightarrow 0^+} h,
	\end{align}
	which implies that $\tilde{h}$ is nonincreasing. Therefore, $f$ is a composition of a convex function $h$ with nonincreasing $\tilde{h}$, and a concave function $g$. Based on the composition rules \cite[Ch.3 \S 3.11]{Boyd-Vandenberghe-04}, $f$ will be a convex function in $R^M$. 
	
	\section{A simplified scenario for two transmissions.}
	\label{F}
	For the scenario of $M=2$ transmissions the optimization problem \eqref{eq:17} simplifies to
	\begin{equation}
		\begin{aligned}
			\label {eq:42}
			& {\text{minimize}}
			& & \mathrm{\frac{1}{2}(\rho_1+\rho_2\epsilon_1)} \\
			& \text{subject to}
			&& \frac{\phi^2}{\rho_1\rho_2} = \epsilon
		\end{aligned}
	\end{equation}
	
	To solve problem \eqref{eq:42} we can utilize the procedure described in Section \ref{4}.A. First, by using \eqref{eq:21} and \eqref{eq:28}  we compute the power terms as functions of $\lambda$. Next, by substituting these expressions for $\rho_1$ and $\rho_2$ in \ref{c4} and solving for $\lambda$ we obtain $\lambda=\frac{\phi}{2\epsilon}\sqrt[3]{\tfrac{1}{4\epsilon}}$. Finally, we compute the values of the power terms as $\rho_1=\phi\sqrt[3]{\tfrac{2}{\epsilon}}$ and $\rho_2=\frac{\phi}{\epsilon}\sqrt[3]{\tfrac{\epsilon}{2}}$. 
	
	For this specific case of $M=2$ we can also utilize the following simpler approach to find the optimal power allocation. First, we rewrite the equality constraint as $\rho_2=\frac{\phi^2}{\rho_1 \epsilon}$. Next, by substituting $\rho_2$ in the objective function of \eqref{eq:42}, we obtain an unconstrained optimization problem with variable $\rho_1$. Then, we compute $\rho_1$ by setting the first derivative of the new objective function to zero as
	\begin{align}
		\label{eq:100}
		\frac{1}{2}-\frac{2\phi^3}{\rho_1^3 \epsilon}=0.
	\end{align}
	By solving \eqref{eq:100} we find $\rho_1=\phi\sqrt[3]{\tfrac{2}{\epsilon}}$, which is same as what we obtained by using the procedure described in Section \ref{4}.A. Then, after substituting the first power term equation in the rewritten equality constraint we compute $\rho_2=\frac{\phi}{\epsilon}\sqrt[3]{\tfrac{\epsilon}{2}}$. 
	
	As pointed out in Section \ref{bbb}, the procedure for CC-HARQ and IR-HARQ will follow a similar pattern.
	
	\section{Proof of Theorem 2.}
	\label{Appendix B}
	From \eqref{eq:32} we can write $\rho_m$ and $\rho_{m+1}$ as:
	\begin{align}
		\sqrt{2^{a(m)} \phi^{b(m)}(M\lambda)^{c(m)}}<\sqrt{2^{a(m+1)} \phi^{b(m+1)}(M\lambda)^{c(m+1)}}.
	\end{align}
	Next, it follows that
	\begin{align}
		&2^{a(m)} \phi^{b(m)}(M\lambda)^{c(m)} < 2^{a(m+1)} \phi^{b(m+1)}(M\lambda)^{c(m+1)},\\
		&\frac{2^{a(m)} \phi^{b(m)}M^{c(m)}}{2^{a(m+1)} \phi^{b(m+1)}M^{c(m+1)}} < \frac{\lambda^{c(m+1)}}{\lambda^{c(m)}}, \\
		&2^{a(m)-a(m+1)}\phi^{b(m)-b(m+1)}M^{c(m)-c(m+1)}\!<\! \lambda^{c(m+1)-c(m)}.
		\label{50}
	\end{align}
	In \eqref{50} the exponent of $\lambda$ is a positive number. To prove that, we can show that $c(m+1)-c(m) > 0$, which leads to
	\begin{align}
		2-2^{-M+m+1}< 2-2^{-M+m+2}.
		\label{51}
	\end{align}
	Inequality holds in \eqref{51} by taking the logarithm on both sides. Next, we substitute the value of $\lambda$ from \eqref{eq:39}. We observe that as $\epsilon \rightarrow 0$, the right-hand side of the inequality in \eqref{50} will tend to infinity. Since all the transformations we have done are equivalent, we can argue that $\rho_m < \rho_{m+1}$.
	
	\section{Proof of Theorem 3.}
	\label{C}
	To obtain the optimal power allocation scheme, we solve problem \eqref{eq:17}. We start our solution by writing the Lagrangian function, which is 
	\begin{align}
		\label{eq:34}
		\mathcal {L} (\rho_m, \mu_m,\lambda) &= \sum_{m=1}^M \rho_m E_{m-1}^{CC-HARQ} + \sum_{m=1}^M \mu_m \rho_m+ \lambda(E_M^{CC-HARQ}-\epsilon),
	\end{align}
	where $E_{m}^{CC-HARQ}$ can be computed from \eqref{eq:33}.
	
	From \ref{c1} we compute the derivative of the Lagrangian function $\mathcal {L} (\rho_m, \mu_m,\lambda)$ with respect to the power $\rho_m$ as
	\begin{align}
		\frac{\partial \mathcal {L} (\rho_m, \mu_m,\lambda)}{\partial \rho_m }&=\Big(\frac{\phi^{m-1}}{(m-1)!\prod_{i=1}^{m-1}\rho_i} - \sum_{i=1}^{M-m}\frac{\rho_{m+i}\phi^{m+i-1}}{(m+i-1)!\rho_m^2\prod_{j=1, j \neq m}^{m+i-1}\rho_j}\Big) \nonumber\\& - \mu_m - \lambda\left(\frac{\phi^M}{M!\rho_m^2 \prod_{i=1, i \neq m}^{M}\rho_i}\right).
		\label{eq:35}
	\end{align}
	
	Following a similar procedure as in the proof of Theorem \ref{th1}, we allow $\mu_m=0$. By taking the derivatives with respect to $\rho_M, \rho_{M-1}, \ldots, \rho_1$ and making some mathematical manipulations we obtain the following structure
	\begin{align}
		\label{j}
		\rho_M&=\sqrt{\phi\lambda},\\
		\label{k}
		\rho_{M-1}&= \sqrt{\frac{2\rho_M\phi}{M-1}}, \\
		\rho_{M-2}&=\sqrt{\frac{2\rho_{M-1}\phi}{M-2}}, \\
		\vdots \nonumber\\
		\rho_{M-m}&=\sqrt{\frac{2\phi\rho_{M-m+1}}{M-m}}.
	\end{align}
	From the structure above, we observe that we can still apply the back-substitution approach and obtain $\rho_m$ as in \eqref{eq:36}.

	Based on \eqref{eq:36}, it is straightforward to show that the power terms are positive. Further, all the power terms are computed as a function of $\lambda$. To find the value of the equality Lagrange multiplier we utilize \ref{c4} and write:
	\begin{align}
		\label{eq:37}
		E_M^{CC-HARQ}=\frac{\phi^M}{M!\prod_{m=1}^M \rho_m}=\epsilon,
	\end{align}
	where $\rho_m$ is calculated as
	\begin{align}
		\label{eq:38}
		\rho_m=\frac{2^a\phi^b\lambda^c}{\sqrt{m}}.
	\end{align}
	In \eqref{eq:38}, the exponents $a(m)$, $b(m)$ and $c(m)$ are defined in \eqref{a(m)}, \eqref{b(m)} and \eqref{c(m)}. Finally, we compute the Lagrangian multiplier $\lambda$ as:
	\begin{align}
		\label{eq:40}
		\lambda=\left( \frac{\phi^{o(M)}}{(M! \epsilon)^{p(M)} \prod_{m=1}^{M-1}\frac{2}{M-m}^{p(M)-p(m)}}\right)^\frac{1}{o(M)},
	\end{align}
	where $o(m)$ and $p(m)$ are given in \eqref{o(M)} and \eqref{p(M)} respectively.
	
	\section{Proof of Lemma 2.}
	\label{D}
	In \cite{Laneman} the author shows that
	\begin{align}
		\lim\limits_{s \rightarrow \infty} s^{m+1} \mathrm{Pr\left[u_m+v_m < R\right]}= \int_{0}^{R} \psi(R-x) \gamma'(x) dx.
		\label{eq:90}
	\end{align}
	where $\psi(R)$ and $\gamma(R)$ are monotone and increasing and integrable functions, $\gamma' (R)$ is integrable and $u_m$ and $v_m$ are independent random variables that satisfy the following conditions
	\begin{align}
		\lim\limits_{s \rightarrow \infty} s \mathrm{Pr\left[u_m < R\right]}= \gamma(R), \nonumber\\
		\lim\limits_{s \rightarrow \infty} s^m \mathrm{Pr\left[v_m < R\right]}= \psi(R).\nonumber
		\label{eq:91}
	\end{align}
	
	In \eqref{eq:50} we set $u_m=\log(1+\rho_m)$. It is straightforward to show that, when the channel gains are Rayleigh distributed 
	\begin{align}
		\mathrm{Pr}\left[u_m < R\right] = 1-e^{\left(\frac{e^R-1}{\rho_m}\right)},
	\end{align}
	and 
	\begin{align}
		\lim\limits_{s \rightarrow \infty} s \mathrm{Pr\left[u_m < R\right]}= e^R-1.
		\label{eq:93}
	\end{align}
	Letting $\psi_0(R)=1$ and $\psi_1 (t)=\gamma(t) e^R-1$, and recursively applying the theorem we obtain
	\begin{align}
		\lim\limits_{s \rightarrow \infty} s^2 \mathrm{Pr\left[\sum_{m=1}^{2}u_m < R\right]}&= \int_{0}^{R} \psi_1(R-x) f'(x) dx \\
		&=R (e^R-1) + 1. 
		\label{eq:94} 
	\end{align}
	The expression computed in \eqref{eq:94} corresponds to $\psi_2 (R)$. By continuing this, we obtain the recursive integral
	\begin{align}
		\lim\limits_{s \rightarrow \infty} s^M \mathrm{Pr\left[\sum_{m=1}^{M}u_m < R\right]}&= \int_{0}^{R} \psi_{M-1}(R-x) \gamma'(x) dx \\
		&=\psi_M (R). 
		\label{eq:95}
	\end{align}
	The recursive integral in \eqref{eq:95} provides each of the terms $\psi_1 (R), \psi_2 (R), \ldots \psi_M (R)$. This integral converges to the series shown in \eqref{eq:98}. Thus, we compute each of the terms $\psi_m (R)$ as
	\begin{align}
		\psi_m (R)= (-1)^m \left( 1+ \sum_{i=0}^{m} \frac{(-1)^i}{(i-1)!} e^R R^{i-1} \right), ~ m=1,\ldots,M.
		\label{eq:98}
	\end{align}
	Finally, by using these results, we can approximate the outage probability as shown in \eqref{eq:41}.
	
	\section{Proof of Theorem 4.}
	\label{E}
	To compute the power allocation algorithm, again we resort to problem \eqref{eq:17}. First, we write the Lagrangian function as 
	\begin{align}
		\label{eq:43}
		\mathcal {L} (\rho_m, \mu_m,\lambda) = \sum_{m=1}^M \rho_m E_{m-1}^{IR-HARQ} + \sum_{m=1}^M \mu_m \rho_m+ \lambda(E_M^{IR-HARQ}-\epsilon),
	\end{align}
	
	By analyzing \ref{c1}, we write the derivative of the Lagrangian function $\mathcal {L} (\rho_m, \mu_m,\lambda)$ with respect to the power $\rho_m$ as
	\begin{align}
		\frac{\partial \mathcal {L} (\rho_m, \mu_m,\lambda)}{\partial \rho_m }&=\left( \frac{\psi_{m-1}}{\prod_{i=1}^{m-1} \rho_i}-\sum_{m=1}^{M-m}\frac{\rho_{m+i}\psi_{m+i-1}}{\rho_m^2 \prod_{i=1, i\neq m}^{m+i-1} \rho_i} \right) -\mu_m-\lambda\left( \frac{\psi_M}{\rho_m^2 \prod_{i=1, i \neq m}^{M}\rho_i} \right)
		\label{eq:44}
	\end{align}
	Similarily as in the repetitive schemes, we relax the non-negativity constraint for the power terms. Next, by derivating with respect to $\rho_M, \rho_{M-1} \ldots \rho_1$ and making some mathematical operations we obtain 
	\begin{align}
		\label{l}
		\rho_M&=\sqrt{\frac{\lambda M \psi_M}{\psi_{M-1}}},\\
		\label{m}
		\rho_{M-1}&= \sqrt{\frac{2\rho_M\psi_{M-1}}{\psi_{M-2}}}, \\
		\rho_{M-2}&=\sqrt{\frac{2\rho_{M-1}\psi_{M-2}}{\psi_{M-3}}}, \\
		\vdots \nonumber\\
		\rho_{1}&=\sqrt{\frac{2\rho_2\psi_1}{\psi_0}}.
	\end{align}
	
	Again, we notice that the back-substitution approach is applicable, and the power term of the $m^{th}$ round can be written as in \eqref{eq:45}.
	
	At this point, we argue that the power terms in \eqref{eq:45} are non-negative. Further, we compute each of the power terms as a function of the equality Lagrangian multiplier, $\lambda$. To derive a closed-form expression for it, we utilize \ref{c4}:
	\begin{align}
		\label{eq:46}
		E_M^{IR-HARQ}=\frac{\psi_M}{M!\prod_{m=1}^M \rho_m}=\epsilon,
	\end{align}
	where the power term in the $m^{th}$ round can be found as
	\begin{align}
		\label{eq:47}
		\rho_m=2^{a(m)} (m \lambda)^{c(m)} \prod_{i=1}^{m} \frac{\psi_i}{\psi_{i-1}}^{d(i)}.
	\end{align}
	In \eqref{eq:47}, $a(m)$ and $c(m)$ can be found from \eqref{a(m)} and \eqref{b(m)}, respectively. Further, the exponent $d(i)$ is found as $d(i)=2^{-i}$. Finally, we can find the value of $\lambda$ as
	\begin{align}
		\label{eq:48}
		\lambda=\left(\frac{\psi_M}{M^{o(M)} \epsilon^{p(M)} 2^{q(M)} \prod_{m=1}^{M} \frac{1}{\psi_{m-1}^{2^{M-m}}}}\right)^\frac{1}{o(M)},
	\end{align}
	where $o(m)$, $p(m)$ and $q(m)$ are given from \eqref{o(M)}, \eqref{p(M)} and \eqref{q(M)} respectively.

	\bibliographystyle{IEEEtran}
	\bibliography{IEEEabrv,Papers1}
	
\end{document}